\documentstyle[12pt,preprint]{aastex}

\shorttitle{Practical Image Masks}
\shortauthors{Kuchner \& Spergel}
\begin{document}

\slugcomment{Revised for ApJ February 26, 2003}
 
\title{Notch Filter Masks: \\
Practical Image Masks for Planet-Finding Coronagraphs}
\author{Marc J. Kuchner\altaffilmark{1}}
\affil{Harvard-Smithsonian Center for Astrophysics \\ Mail Stop 20, 60 Garden St., Cambridge, MA 02138}
\altaffiltext{1}{Michelson Postdoctoral Fellow}
\email{mkuchner@cfa.harvard.edu}

\author{David N. Spergel}
\affil{Princeton University Observatory \\ Peyton Hall, Princeton, NJ 08544}
\email{dns@astro.princeton.edu}

\begin{abstract}

An ideal coronagraph with a band-limited image mask can 
efficiently image off-axis sources while removing identically
all of the light from an on-axis source.  
However, strict mask construction tolerances limit the utility of
this technique for directly imaging extrasolar terrestrial
planets.  We present a variation on the basic band-limited
mask design---a family of ``notch filter'' masks---that
mitigates this problem.  These robust and trivially achromatic
masks can be easily manufactured by cutting holes in
opaque material.

\end{abstract}

\keywords{astrobiology --- circumstellar matter --- 
instrumentation: adaptive optics --- planetary systems}

\section{INTRODUCTION}

Direct optical imaging of nearby stars
has emerged as a potentially viable method for detecting
extrasolar terrestrial planets,
buoyed by new techniques for controlling
diffracted and scattered 
light in high-dynamic-range space telescopes
(see, e.g., the review by Kuchner \& Spergel 2003).
These techniques boost a telescope's ability to separate a planet's
light from the light of its host star.
At optical wavelengths, the Sun outshines the Earth by a
factor of nearly $10^{10}$; this contrast ratio is $\sim 10^{3}$ times
larger than the contrast ratio in the mid-infrared
\citep{tpf,desm01}.  But to offset the higher dynamic range requirements
of visible-light planet finding, optical techniques offer freedom from
large, multiple-telescope arrays \citep{wool03}, cryogenic optics, and
background light from zodiacal and exozodiacal dust \citep{kuch00},
while providing access to $O_2$ and $O_3$ biomarkers
\citep{trau01,desm01}, surface features \citep{ford01},
the total atmospheric column density \citep{trau03}, and even potentially
the ``red edge'' signal from terrestrial vegetation \citep{wool02}.

Of the obstacles to
achieving the necessary dynamic range in a single-dish optical telescope, 
the diffracted light background appears relatively manageable.
For example, maintaining the scattered light background at the level of the
expected signal from the planet poses a greater challenge; this task
requires a r.m.s. wavefront accuracy of $\lesssim 1$~\AA~\citep{kuch02, trau02a}
over the critical spatial frequencies.
However, techniques for managing the diffracted light may dictate
the general design of a planet-finding telescope and the
planet search and characterization strategy.

Optical techniques for controlling diffracted light in
planet-imaging telescopes have centered on two main designs: 
specially shaped and/or apodized pupils
\citep{sper01,nise01,kasd01,debe02,kasd03} and classical
coronagraphs \citep{lyot39, naka94, stah95, malb95, kuch02}.
Shaped and apodized pupils produce a point spread function
whose diffraction wings are suppressed in some regions of the
image plane.  A classical coronagraph 
explicitly removes the on-axis light from the optical
train by reflecting or absorbing most of it with an image
mask and diffracting the remainder onto an opaque
Lyot stop.

Recently, \citet{kuch02} showed that a classical coronagraph
performs best with a ``band-limited'' image mask.  
Different band-limited masks offer high performance for planet
searching or planet characterization.  For planet characterization,
the $\sin^2$ amplitude transmissivity mask ($\sin^4$ intensity transmissivity)
introduced in \citet{kuch02} can achieve 80\% throughput for a
planet at 4$\lambda/D$.  With this high throughput, a 10 m by 4 m telescope
can detect a planetary biomarker in $\sim$1/3 of the time needed by
alternative designs (e.g., an 8 m square apodized aperture).  A
band-limited mask of the form $1-{\rm sinc}$ (see Table~1) has both
excellent throughput and large search area.  With any 
band-limited mask, an ideal coronagraph eliminates identically all
of the on-axis light, though pointing errors and the stellar size contribute
to a finite leakage \citep{kuch02}.  A band-limited mask can
operate with a pupil of any shape as long as it has uniform
transmissivity.

But because they interact with focused starlight, all coronagraphic
image masks face severe construction
tolerances.  Errors in the mask intensity transmissivity
of $\sim 10^{-9}$ on scales of $\lambda/D$
near the center of the mask can scatter enough light into
the field of view to scuttle a planet search \citep{kuch02}.
Painting a graded-transmissivity mask requires a steady hand!
This requirement has cast the classical coronagraph in an unfavorable
light, despite its potential high performance and flexibility.

In this paper, we offer a way around this pitfall of
classical coronagraphy: an easy-to-manufacture class of image
masks.   We illustrate a family of binary image masks which
offer a savings in construction tolerances of $\sim5$ orders of
magnitude compared to graded image masks, analogous to the advantage
of using binary rather than graded pupil masks \citep{sper01}.
These ``notch filter'' masks offer the same planet
search and characterization advantages as ideal band-limited masks,
providing a robust, practical means of controlling diffracted light in
a planet-finding coronagraph.

\section{BAND-LIMITED MASKS}
\label{sec:bandlimitedmasks}

We begin by reviewing the theory of band-limited image masks.  We retain
the notation of \citet{kuch02}; image plane quantities have hats
and pupil plane quantities do not.

\begin{figure}
\epsscale{0.85}
\plotone{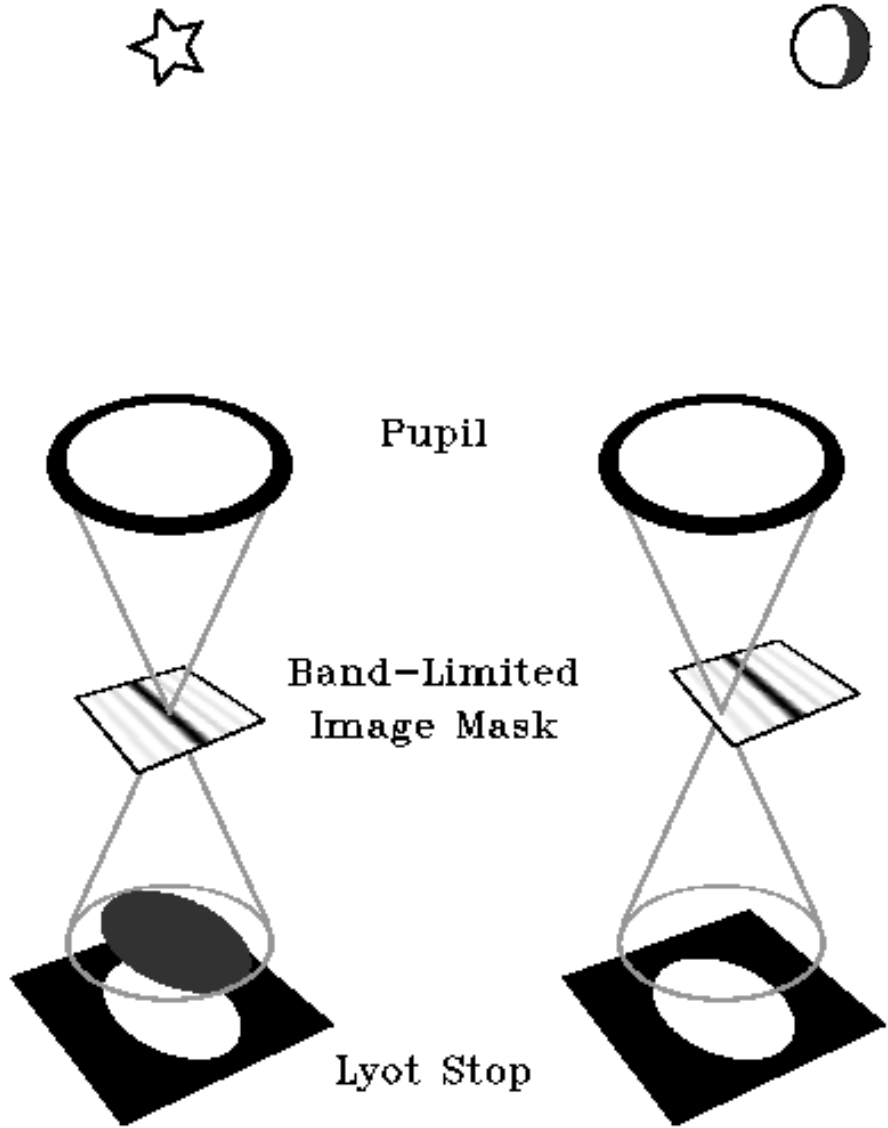}
\caption{Cartoon of a coronagraph with a band-limited image
mask.  The image mask diffracts on-axis starlight to a region
restricted to the edges of the pupil plane, where a Lyot stop 
blocks it.  Off-axis light from a planet diffracts all
around the pupil plane, and through the center of the Lyot stop.
\label{fig:cartoon}}
\end{figure}

Figure~\ref{fig:cartoon} illustrates schematically how a
coronagraph works; light passes through the pupil and 
converges on an image mask,
then the pupil is re-imaged onto a Lyot stop.
Starlight focused on the center of the image mask
diffracts to the pupil edges, where the Lyot stop can block it,
as shown on the left of the figure.
Light from an off-axis planet diffracts all around the second
pupil plane, as shown on the right of the figure,
and largely passes through the Lyot stop.


A band-limited mask has a transmission function chosen
to diffract all the light from an on-axis source to angles
within $\epsilon D/(2 \lambda)$ of the edges of the pupil, as shown in
Figure~\ref{fig:cartoon}, so that a well-chosen Lyot stop can block identically
all of that diffracted light.  Such an image mask typically
consists of a series of dark rings or stripes.
The parameter, $\epsilon$, is the bandwidth of the mask.

A mask can be described by an amplitude transmissivity, $\hat M(x,y)$,
and intensity transmissivity $|\hat M(x,y)|^2$ where $x$ and $y$ are
cartesian coordinates in the image plane.
Image masks are generally opaque ($\hat M=0$) in the center ($x=y=0$)
and close to transparent ($\hat M \approx 1$) away from the center, in the search area.
To understand the need for band-limited pupil masks,
we must examine the Fourier transform of $\hat M(x,y)$, given by
\begin{equation}
M(u,v)=\int \! \! \int \hat M(x,y) \, e^{-2 \pi i (ux+vy)} \, dx \,dy
\end{equation}
The amplitude transmissivity of a completely transparent mask has
only one Fourier component, at zero frequency, i.e.
$M(u,v) = \delta(u,v)$.

\begin{figure}
\epsscale{0.85}
\plotone{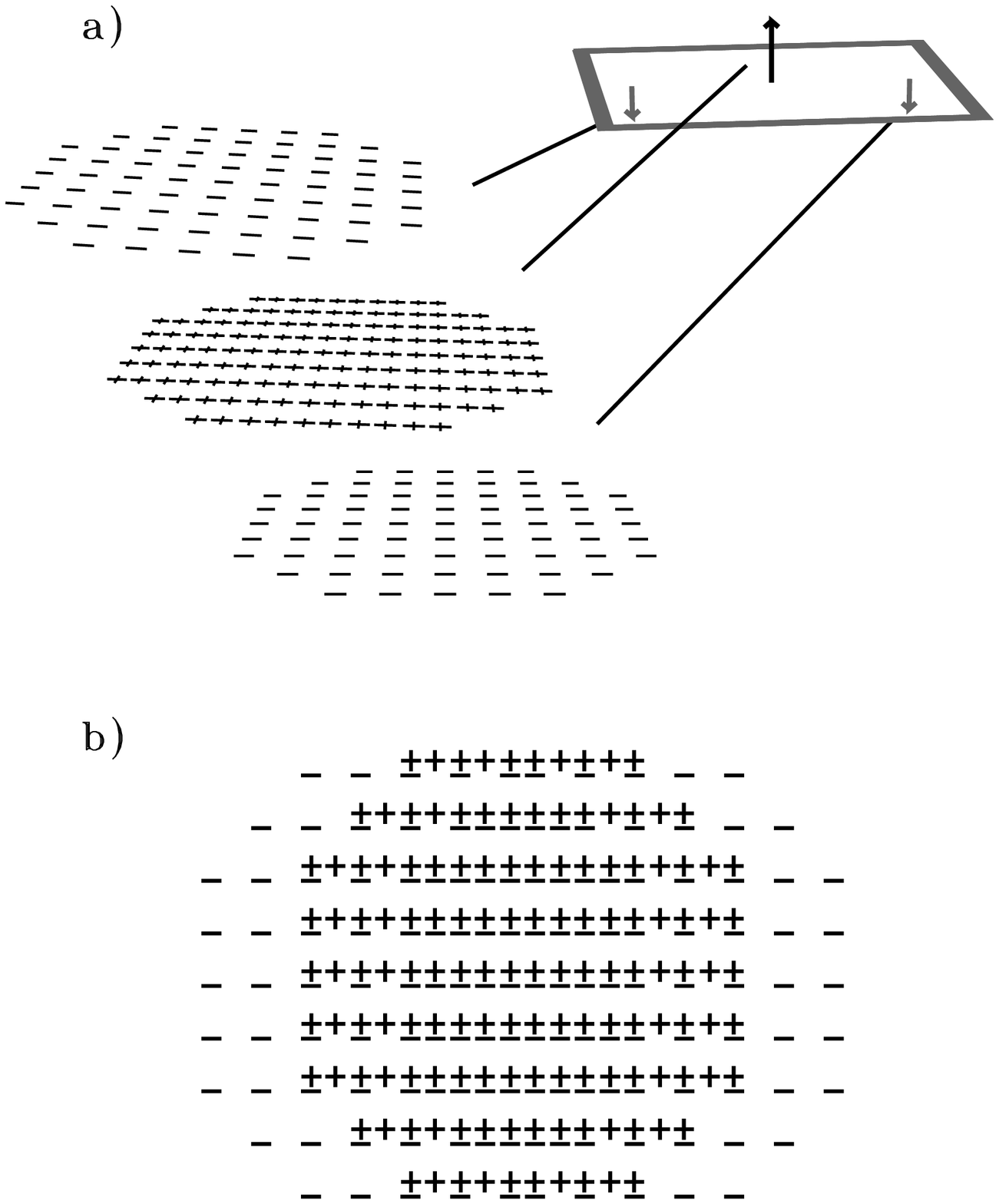}
\caption{Convolving the pupil field of an on-axis source
with the Fourier transform of $\sin^2$.  a)  For each
delta function in the Fourier transform, there is one
weighted copy of the field, a virtual pupil.  b)  The field in the
second pupil plane is the sum of the fields of the
virtual pupils.  The fields cancel to zero in the center where
there are equal densities of plus and minus signs.
\label{fig:phasor}}
\end{figure}

Figure~\ref{fig:phasor} illustrates the
operation of a mask with one cosine
component besides the zero-frequency component, the $\sin^2$ mask
($\sin^4$ intensity transmissivity) described in \citet{kuch02}.  
The Fourier transform of the amplitude transmissivity of this
mask consists of three delta functions:
\begin{equation}
M(u,v)=
-{1 \over 4} \delta(u+\epsilon D/(2 \lambda),v)
+{1 \over 2} \delta(u,v)
-{1 \over 4} \delta(u-\epsilon D/(2 \lambda),v)
\label{eq:threedeltas}
\end{equation}
This mask is the
simplest example of a band-limited mask.


The amplitude transmissivity, $\hat M(x,y)$, multiplies the field amplitude
in the image plane.  In the pupil plane, on the other side of a Fourier
transform, this multiplication becomes a convolution.
Figure~\ref{fig:phasor}a illustrates the convolution of the
amplitude of the pupil field of an monochromatic on-axis source
and the function, $M(u,v)$, given in Equation~\ref{eq:threedeltas}.

In the convolution, each $\delta$-function from Equation~\ref{eq:threedeltas}
generates a weighted copy of the pupil field---a virtual pupil.
We represent each copy of the field
as a circle filled with $+$ signs or $-$ signs.  The circular
shape represents a circular aperture, though any aperture shape will do.
Since the central $\delta$-function has twice the weighting of
the other $\delta$-functions, the $+$ signs have twice the density of
the $-$ signs in Figure~\ref{fig:phasor}a.

Figure~\ref{fig:phasor}b depicts the field in the second pupil plane, 
the sum of the three virtual pupil fields shown in
Figure~\ref{fig:phasor}a.  In the center of
Figure~\ref{fig:phasor}b, the densities of
$+$ and $-$ signs are equal; for every $+$ sign, there is a $-$
sign.  In this region, the fields cancel to zero.  Elsewhere
the fields do not cancel.  The next optical element in the
coronagraph beam train is a Lyot stop, which transmits light in the
center of the pupil plane, but blocks the regions where the fields
do not cancel.

A given Lyot stop blocks the light diffracted by a range
of Fourier components.  If the the Lyot stop blocks a
fraction, $\epsilon$, of the pupil radius at the pupil
edges, it will block the diffracted
light from all spatial frequency components in the mask with
spatial frequency $|u| < \epsilon D/(2 \lambda)$,
where $D$ is the telescope diameter, and $\lambda$ is the
wavelength.  One can create a mask which contains any or all of the
cosine Fourier components at these low frequencies which
the Lyot stop will still match; this family of masks
which has power in only a limited range of low spatial frequencies
is the family of band-limited masks.  We can use $\epsilon$
to refer to the bandwidth of a Lyot stop or the bandwidth of an
image mask matched to that Lyot stop.  

Likewise, a given ideal Lyot stop and
mask combination can work at a range of wavelengths.
The bandwidth of a given image mask is proportional to $\lambda$,
but the bandwidth of a given Lyot stop is independent of $\lambda$.
Therefore, a given Lyot stop/image mask combination
will work at all wavelengths shorter than the wavelength for
which it was designed.  However, it can only have optimum
throughput at one wavelength.

\citet{kuch02} display a variety of one-dimensional
band-limited mask amplitude transmissivity functions.
A useful compromise between search area and throughput is
$\hat M(x)=N \left(1- {\rm sinc}(\pi x \epsilon D/\lambda)\right)$,
where ${\rm sinc}\,{x}=\sin(x)/x$, and $1-1/N$ is the minimum value of
${\rm sinc}\,{x}$ (i.e., $N = 0.82153497637881...$).
The throughput of a Lyot stop matched to a one-dimensional
mask function is roughly $1-\epsilon$. 
Band-limited masks with additional Fourier components in the $v$ direction
are also possible, though to use these masks, one must
stop the top and bottom of the pupil plane as well as the left
and right.  The throughput of such a Lyot stop
is roughly $(1-\epsilon)^2$.

At the request of NASA, a university-industry team associated
with Ball Aerospace and Technologies Corporation studied a
design for a space-based
visible-light planet finding telescope using a single
4~m by 10~m elliptical primary mirror.  This team estimated that
with a classical coronagraph using a Gaussian image mask,
the design could detect an Earth twin orbiting a G2 V star
at a distance of 10 pc in 0.86 hours, including time for 2 rotations
of the image plane \citep{beic02}.  Once the location of the
planet was known, a water band in the planet's atmosphere
could be detected spectroscopically
in 0.14 days, and an $O_2$ band could be detected in 0.8 days.
If a single band-limited image mask of the form
$\hat M(x)=1-{\rm sinc}^2{x}$ were used instead of a Gaussian mask,
the Lyot stop could be substantially widened,
increasing the throughput, and the detection and characterization times
would be reduced by a factor of roughly 0.7, to 0.6 hours for detection,
0.1 days for $H_2O$, and 0.6 days for $O_2$.

Ideally, a band-limited mask combined with a Lyot stop
completely blocks all on-axis starlight from reaching the second
image plane, and attenuates off axis starlight to an easily 
manageable level.   However, \citet{kuch02} discuss two significant
limitations on the band-limited mask performance: pointing
errors and errors in mask construction. 
None of the time estimates in the Ball report
accounts for either of these errors, which
affect all masks, band-limited or not.

A mildly apodized Lyot stop can compensate for the
leakage due to pointing errors \citep{kuch02}. 
Apodizing the Lyot stop carries a throughput penalty, but
even with this loss, the
ideal classical coronagraph outfitted with a
choice of band-limited masks remains by far the fastest of the
idealized designs described in the Ball report for planet detection.
Mask errors are more serious; all graded image masks suffer
from impractically tight construction tolerances.
We will show how to dramatically loosen
the construction requirements by building binary masks.

\section{NOTCH FILTER MASKS}
\label{sec:notchfiltermasks}

\begin{figure}
\epsscale{0.9}
\plotone{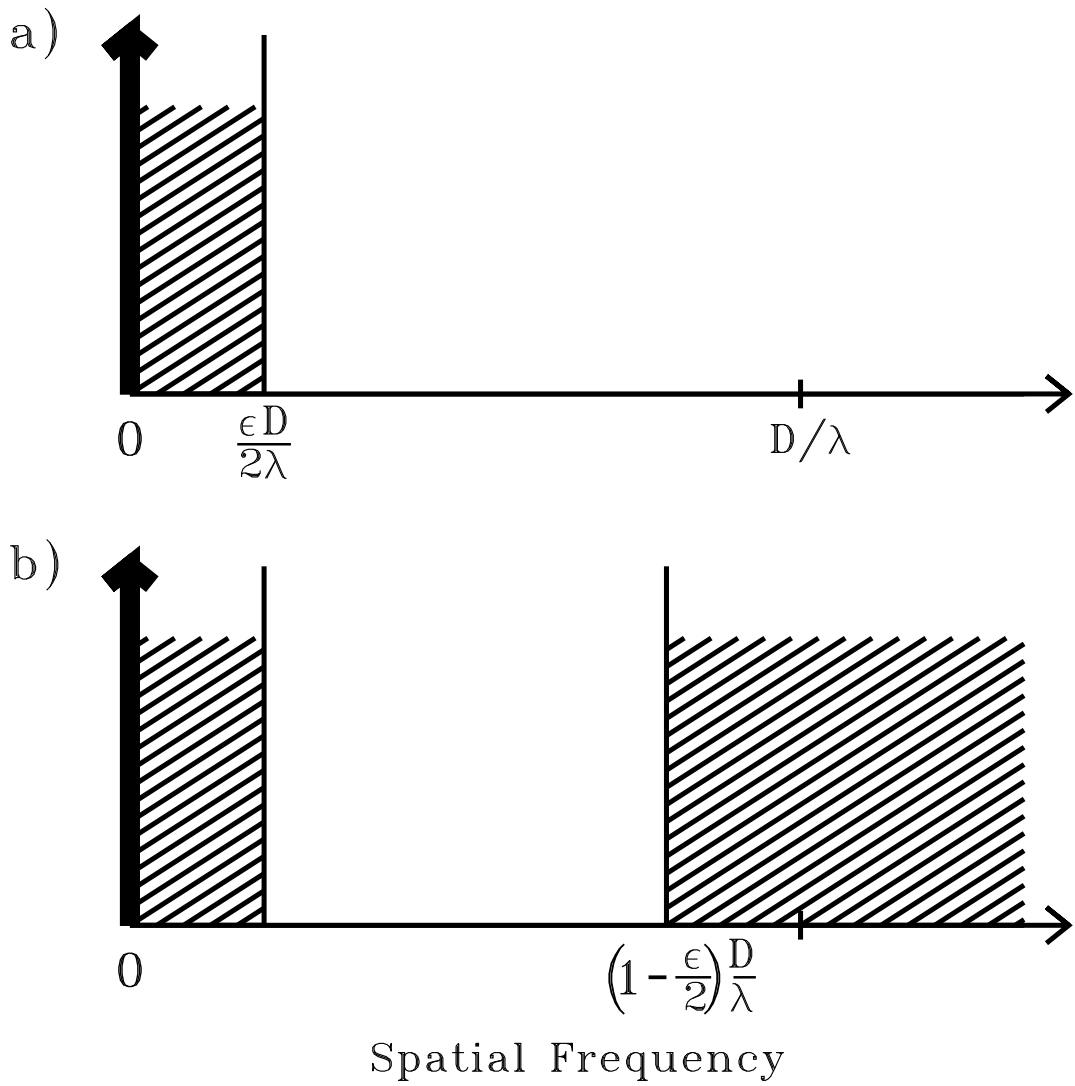}
\caption{Power spectra of a band-limited mask (a) and a
notch filter mask (b).
The mask functions may have power at spatial frequencies
indicated by the hatched regions.  
\label{fig:powerspectrum}}
\end{figure}

To build a binary mask that retains the advantages of band-limited
masks we will need to use more of the available function
space for mask design.
Section~\ref{sec:bandlimitedmasks} reviewed the utility of
masks whose Fourier components are limited to spatial
frequencies $|u| < \epsilon D/(2 \lambda)$.
However, there is another range of spatial frequencies
available for mask design: as \citet{kuch02} described in their
discussion of mask errors, high spatial frequency terms
that diffract light well outside the opening in the Lyot stop do
not affect the performance of a mask.  We can add
high spatial-frequency terms, with $|u| >  (1- \epsilon/2) (D/\lambda)$,
to the mask amplitude transmissivity function without altering the
light admitted through the coronagraph as long as 
\begin{equation}
\int_{-\epsilon D/(2 \lambda)}^{\epsilon D/(2 \lambda)} M(u) \,du= 0.
\label{eq:cancellation}
\end{equation}
Figure~\ref{fig:powerspectrum} shows that the 
spatial frequency response of a general image mask which can
completely suppress on-axis light resembles
the spectral response of a notch filter.


We can use the degrees of freedom available at high spatial frequencies
to design masks which are relatively easy to
construct to the necessary tolerances.  
For example, the transmissivity of a band-limited
mask is analytic, so it 
can not have a constant value over any finite region.
However, the transmissivity of a notch filter
mask need not obey this restriction.  The remainder of this paper
will be a discussion of notch filter masks that take advantage of
this opportunity.

\section{ONE-DIMENSIONAL MASK FUNCTIONS}

Sampling a function forces its Fourier transform to be periodic.
We can harness this aliasing effect to generate useful notch filter mask functions.
We will illustrate this principle first by considering mask functions
of one variable only.  Such a mask function can be realized as a
striped mask as shown in
\citet{kuch02}.  These functions can also be used as parts to
construct two-dimensional masks.

Throughout our discussion, $\hat M_{BL}(x)$ will be a function which
can serve as the amplitude transmissivity of a band-limited mask:
$0 \le \hat M_{BL}(x) \le 1$, $\hat M_{BL}(0)=0$, and the Fourier transform, $M_{BL}(u)$,
of this function only has power at spatial frequencies
$u < \epsilon D/(2 \lambda)$.  Such a function automatically satisfies
Equation~\ref{eq:cancellation}.
We shall use $\hat M_{BL}(x)$ to create notch filter functions,
$\hat M_{notch}(x)$, that mimic $\hat M_{BL}(x)$ at low spatial
frequencies.

\subsection{Sampling}
\label{sec:sampling}

Actual image masks are constructed using finite-sized
tools offering limited contrast and spatial resolution.  We can design
notch filter masks with this contraint in mind.
Multiply $\hat M_{BL}(x)$ by a comb filter with spacing $\Delta x$
to get a sampled version of $\hat M_{BL}(x)$, 
and convolve the result with a function $\hat P(x)$,
to get
\begin{mathletters}
\begin{eqnarray}
\hat M_{sampled}(x)
&=&\hat P(x) * \left(\hat M_{BL}(x) \,{\Delta x} \sum_n \delta(x-(n+\zeta)\Delta x)\right)  \\
M_{sampled}(u)
&=&P(u) \left(M_{BL}(u) * \sum_n \delta(u-n/\Delta x) e^{-2 \pi i u \zeta \Delta x} \right),
\label{eq:tsampled}
\end{eqnarray}
\end{mathletters}
where $n$ ranges over all integers and $*$ indicates convolution.  For generality, we have allowed the sampling points to be offset from
the mask center by a fraction $\zeta$ of $\Delta x$.
The kernel, $\hat P(x)$, can represent the ``beam'' of a
nanofabrication tool.  It should be normalized so that
$\int_{-\infty}^{\infty} \hat P(x) \, dx = 1$, and 
$\hat P(x)$ must be everywhere $\le 1/(\Delta x)$, so $\hat M_{sampled}(x)$ remains $\le 1$.

This function we have created, $M_{sampled}(u)$, only has power at
$|u-n/{\Delta x}| <  \epsilon/2$.  Its power spectrum resembles
Figure~\ref{fig:powerspectrum} as long as the spacing between
the samples satisfies the requirement
\begin{equation}
\Delta x \le \lambda/D.
\label{eq:deltax}
\end{equation}
If $\zeta \ne 0$, then $M_{sampled}(u)$ typically
has an imaginary component.  However, if
Equation~\ref{eq:deltax} holds,
$M_{sampled}(u)$ is always purely real at
low frequencies ($u < \epsilon D/(2\lambda)$).

In general, $M_{sampled}(u)$ does not
match $M_{BL}(u)$ at low frequencies, because the envelope function,
$P(u)$, in Equation~\ref{eq:tsampled}
does not generally equal unity over the whole bandwidth of $\hat M_{BL}(x)$.  Rather, the
envelope function tends to cause $M_{sampled}(u)$ to violate
Equation~\ref{eq:cancellation}.  However, we can often correct for this
effect and create a function, $M_{notch}(u)$, which satisfies
Equation~\ref{eq:cancellation} by subtracting a constant, $\hat M_0$, from
$\hat M_{sampled}(x)$.  I.e., 
\begin{equation}
\hat M_{notch}=\hat M_{sampled} - \hat M_0,
\end{equation}
where
\begin{equation}
\hat M_0=\int_{-\epsilon D/(2 \lambda)}^{\epsilon D/(2 \lambda)} M_{sampled}(u) \, du 
=\int_{-\infty}^{\infty} M_{BL}(u) P(u) \, du 
=\left. \hat M_{BL}(x) * \hat P(x) \right|_{x=0}.
\label{eq:m0}
\end{equation}

To use this technique, we must not sample $\hat M_{BL}(x)$ where $\hat M_{BL}(x)=0$,
or else we will end up specifying a mask with $\hat M_{notch}(x) < 0$.
Specifically, $|\zeta|$ must be greater than some minimum value, $\zeta_0$,
given by the condition that $\hat M_{BL}(\zeta_0 \Delta x)=\hat M_0$.
We may symmetrize the mask if we desire by forming
a combination such as $(\hat M_{notch}(x) + \hat M'_{notch}(x))/2$, or
$\hat M_{notch}(x)\hat M'_{notch}(x)$ where for $\hat M'_{notch}(x)$, $-\zeta$ has
been substituted for $\zeta$.  The latter combination has twice
the bandwidth of the former.

For example, if we choose $\Delta x=\lambda/D$, and
$\hat P(x)=({D}/{\lambda}) \Pi(x D/\lambda)$, where $\Pi(x)$ is a
tophat function,
\begin{equation}
\Pi(x) = 
\left\{
\begin{array}{ll}
1  & \mbox{where $-1/2<  x < 1/2$ } \\
0  & \mbox{elsewhere},
\end{array}
\right.
\end{equation}
then $P(u)= {\rm sinc}(\pi u \lambda/D)$, and
our sampling algorithm generates a mask resembling
a histogram plot.  A graded version of this mask would
consist of stripes of different uniform shadings.
Choosing $\zeta=\zeta_0$ will generate a striped mask whose
darkest stripe is perfectly opaque. 
Choosing $\zeta_0 > \zeta \le 0.5$ will generate a striped mask which
never becomes perfectly opaque, a potentially useful trick
since fabricating graded masks
with high optical densities can be a challenge \citep{wils02}.

If $\hat M_{BL}(x)=\sin^2(\pi x \epsilon D/(2\lambda))$, then
Equation~\ref{eq:m0} tells us that for this mask,
$\hat M_0=(1/2)(1-{\rm sinc}(\pi \epsilon/2))$,
and $\zeta_0$ is given by the condition
$\hat M_{BL}(\zeta_0 \lambda/D) = \hat M_0 = \sin^2(\pi \zeta_0 \epsilon/2)$.
The trick probably only works for the $\sin^2$ mask when $\epsilon=1/n$,
since $\sin^2$ has so many zeros.
Table 1 lists $\hat M_0$ and $\zeta_0$ for several other masks,
given a tophat kernel.

\begin{table}[h]
     \caption{Sampled Mask Parameters for $\hat P(x)=({D}/{\lambda}) \Pi(x D/\lambda)$}
      \medskip
\begin{tabular}{llccc} \hline\hline
$\hat M_{BL}(x)$                                            & $\hat M_0(\epsilon)$                                                                 & $\epsilon$ & $\hat M_0$          & $\zeta_0$    \\ \hline\hline
$\sin^2{{\pi x \epsilon D} \over {2\lambda}}$                  & ${1 \over 2}\left[1-{\rm sinc}(\pi \epsilon/2)\right]$                                & 0.2        & 0.0081842 & 0.2883579 \\
                                                       &                                                                                 & 0.4        & 0.0322554 & 0.2873981 \\
$N\left[1-{\rm sinc}{{\pi x \epsilon D} \over {\lambda}}\right]$ & ${N \over 2}\left[1-{2 \over \epsilon}\int_0^{\epsilon/2} {\rm sinc}(\pi u') \, du'\right]$ & 0.2        & 0.0022456 & 0.2039059 \\
                                                       &                                                                                 & 0.4        & 0.0089032 & 0.2032511 \\
$1-\left({\rm sinc}{{\pi x \epsilon D} \over {2\lambda}}\right)^2$        & ${1 \over 2}\left[1-{4 \over \epsilon}\int_0^{\epsilon/2} (1-{{2u'} \over {\epsilon}}) {\rm sinc}(\pi u') \, du'\right]$ & 0.2    & 0.0013681 & 0.1019893\\
                                                       &                                                                                 & 0.4        & 0.0054400  &  0.1017713\\
\hline\hline
\end{tabular}
\tablenotetext{}{$N \le 0.82153497637881...$}
\label{tab:table1}
\end{table}

\begin{figure}
\epsscale{0.9}
\plotone{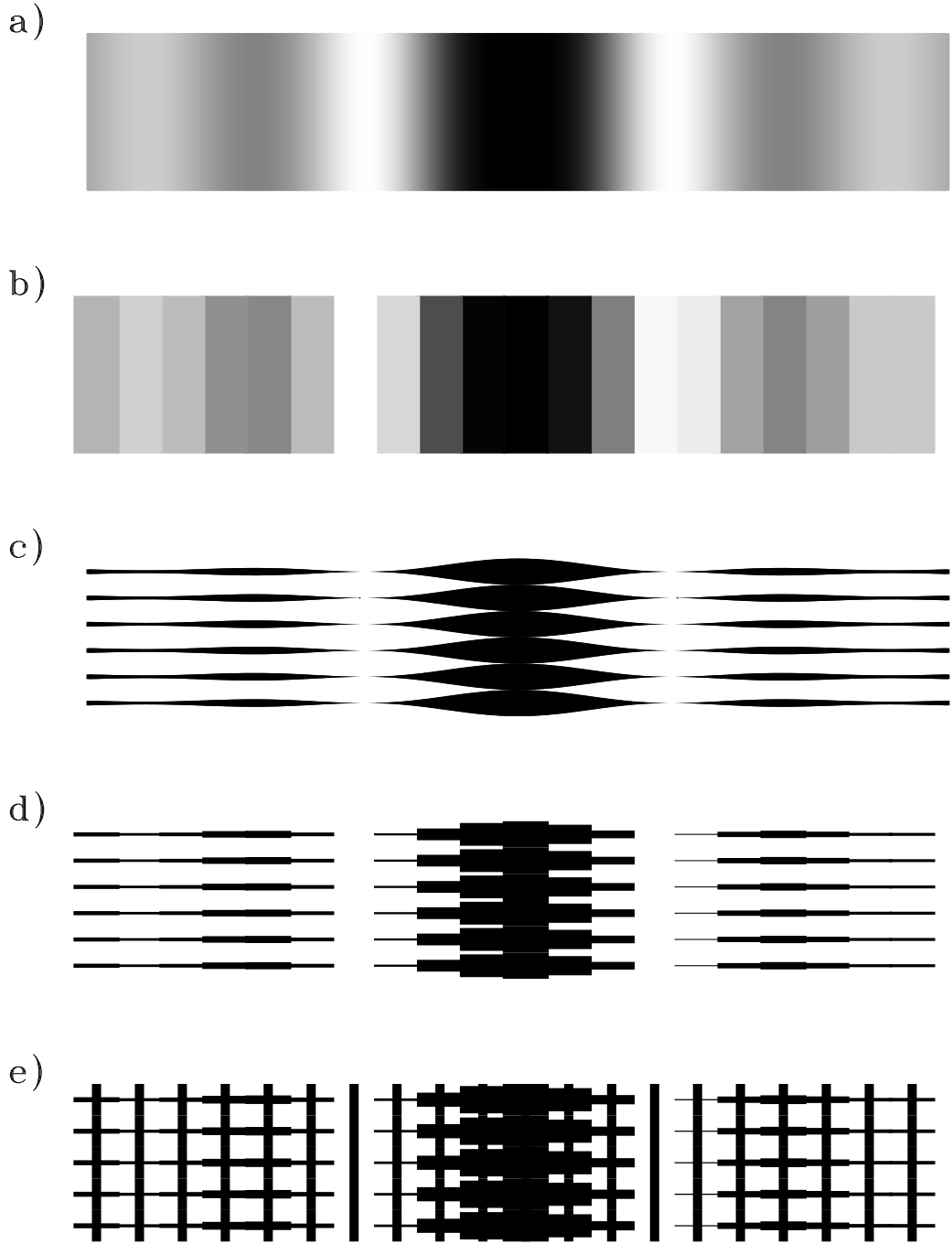}
\caption{Close-up view of the center of a one-dimensional linear
graded band-limited mask (a), and four notch-filter versions of this design
(b,c, d, and e).
\label{fig:linearmasks}}
\end{figure}

Figure~\ref{fig:linearmasks} shows five examples of notch filter masks
all of which are different versions of the same basic $1-{\rm sinc}$ mask.
Figure~\ref{fig:linearmasks}a is a simple band-limited mask
with no additional high-frequency components.
Figure~\ref{fig:linearmasks}b shows a version of this mask sampled as
described above with a tophat kernel of width $\lambda/D$.
This kernel is the narrowest one that works with
$\Delta x = \lambda/D$; narrower kernels require finer samplings.


\section{BINARY MASKS}
\label{sec:binarymasks}

In two dimensions, we can use the additional degrees of
freedom afforded by the high-frequency terms in a notch filter function
to generate a completely binary mask, i.e., a mask
which everywhere satisfies $\hat M_{binary}(x,y)=0$ or $\hat M_{binary}(x,y)=1$.
Such a mask can be constructed entirely out of material which
is highly opaque, like metal foil.

\subsection{Linear Binary Masks}

Let
\begin{equation}
\hat M_{stripe}(x,y)=\left\{
\begin{array}{ll}
0  & \mbox{where $|y| <   \hat M_{notch}(x)\lambda/(2D)$ } \\
1  & \mbox{elsewhere}
\end{array}
\right. 
\end{equation}
and
\begin{equation}
\hat M_{binary}(x,y)= \left(\sum_{n} \delta(y-n \lambda/D)\right) * \hat M_{stripe}(x,y).
\label{eq:tbinary}
\end{equation}
The Fourier transform of this convolution is a product: 
\begin{eqnarray}
\hbox{\hskip -0.25 in}M_{binary}(u,v)
&=&\left({{D}\over{\lambda}}\sum_{n} \delta(v-n D/\lambda)\right) \int\!\!\int \hat M_{stripe}(x,y) e^{- 2 \pi i(ux+vy)} \, dx \, dy  \nonumber \\
&=&\left(\sum_{n} \delta(v-n D/\lambda)\right) \int {\rm sinc}(\pi v (\lambda/D) \hat M_{notch}(x)) \hat M_{notch}(x) e^{-2 \pi iux}\, dx.
\end{eqnarray}
At low and mid-spatial frequencies, only the $v=0$ term contributes, and
\begin{displaymath}
M_{binary}(u,v) = M_{notch}(u) \approx M_{BL}(u) \qquad \mbox{for $u,v < \epsilon D/(2 \lambda)$}.
\end{displaymath}

Though $\hat M_{notch}(x)=1$ in some places,
it is possible to multiply $\hat M_{notch}(x)$ by a positive real constant,
less than 1, to allow for a mask substrate that is not perfectly transparent
or reflective, or to guarantee that the metal strips maintain at least
a minimum width, at a small cost in throughput.

If we use a sampled mask function for $\hat M_{notch}(x)$, the binary mask
may be constructed entirely from opaque rectangles of varying lengths,
for example, generating a ``manhattan'' pattern for simple nanofabrication.
Figures~\ref{fig:linearmasks}c, d and e show examples of binary
masks which mimic the $1-{\rm sinc}$ mask.
Figures~\ref{fig:linearmasks}d and e are binary sampled masks.

\subsection{Circular Binary Masks}
\label{sec:circular}

We recommend using a linear mask for the following reasons:
1) Linear masks have bandwidth in one direction only,
so they generally have the best throughput.
2) If one region of the mask deteriorates, the mask may simply
be translated so that the starlight falls on a new region.
3) Errors in the uniformity of the wavefront in the
direction perpendicular to the image mask cancel out in the
Lyot plane; for example, the telescope need only be pointed
accurately in one direction.
4) It may be possible to use a carefully oriented
linear mask to block the light from a binary star.

\begin{figure}
\epsscale{0.8}
\plotone{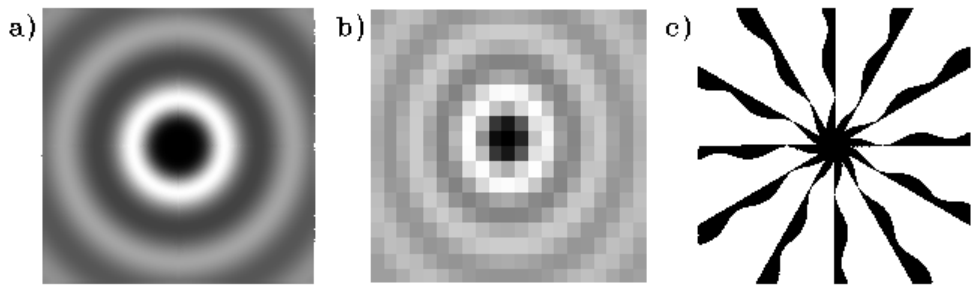}
\figcaption{Close-up view of the center of a one-dimensional radial
band-limited mask (a), and two equivalent notch-filter versions (b and c).
\label{fig:radialmasks}}
\end{figure}

However, circularly
symmetric image masks can provide slightly more search
area than linear image masks, so we discuss them here.
Figure~\ref{fig:radialmasks}a shows the center of a graded band-limited mask
of the form $\hat M_{BL}(r)=N(1-{\rm sinc}\,{\pi r \epsilon D/\lambda})$.
Figure~\ref{fig:radialmasks}b shows a sampled version of this mask,
where, necessarily, the sampling has been performed in two-dimensions.
Creating this sampled mask requires following the same procedure illustrated
in Section~\ref{sec:sampling} to guarantee that the mask satisfies
Equation~\ref{eq:cancellation}.   The sample points are shifted by a
fraction of $\lambda/D$ in some direction, and a constant is subtracted
from the mask amplitude transmissivity.

We can also replace an azimuthally symmetric transmission function, $\hat M(r)$,
with a discrete K-fold symmetric ``star'' mask.  First, choose
a 1-dimensional band-limited function, $\hat M(r)_{BL}$ or a notch filter
version, $\hat M_{notch}(r)$.  Then let
\begin{equation}
\hat M_{binary}(r, \theta) = 
\left\{
\begin{array}{ll}
1  & \mbox{where ${\rm mod}(K \theta/2\pi,1) < \hat M_{notch}(r)$ } \\
0  & \mbox{elsewhere}
\end{array}
\right. 
\end{equation}
The Fourier transform of this function (see, e.g., Jackson (1975), p131 [problem 3.14])  is
\begin{eqnarray}
M_{binary}(q,\phi)=
\lefteqn{\int_0^{\infty} \hat M(r) J_0(qr) r \,dr}  \nonumber \\
&&+ \sum_{m=-\infty}^{m=\infty} 
i^m \exp(imK\phi) \int_0^{\infty} J_{mK}(qr) \sin(m\hat M_{notch}(r)) r \, dr
\label{eq:starmaskft}
\end{eqnarray}
where $q$ and $\phi$ are the radial and angular polar coordinates
in the pupil plane, and $J_m$ is the Bessel function of order $m$.
Figure~\ref{fig:radialmasks}c shows an example of such a binary
star mask.  

For a truely band-limited mask, the radial integrals in
Equation~\ref{eq:starmaskft} should be evaluated over a range from 0 to infinity.
However, as \citet{kuch02} discussed, a mask truncated at a radius of
say, $r=100 \lambda/D$, can serve more than adequately as an
approximation to a band-limited mask.  Moreover, the mask illumination
falls off rapidly with $r$, so deviations from an ideal mask are
inconsequential outside some radius $r_{max}$, which is likely to be much less
than $100 \lambda/D$.

If we consider the mask to be truncated at $r=r_{max}$, 
the high frequency terms are significant only for high $q$.
Since $J_{mK}(qr) \approx (qr)^{mK}/(2^{mK} (mK)!)$ for $qr < mK$, 
the higher order terms must have absolute values less
than $(Dr_{max}/(4\lambda))^{K}/K!$ inside the pupil ($q \le D/(2\lambda)$);
with enough points in the star, these terms are all small.
For example, if $r_{max} = 10 \lambda/D$, $\hat M$ will be less than
$10^{-5}$ interior to the Lyot stop for $K \ge 14$; this level
suffices to allow a coronagraph to suppress the intensity of an on-axis
source in the final image plane by a factor of $10^{-10}$.
 
\subsection{Combining Notch Filter Masks}

In general, the product of two notch filter mask functions is not
a notch filter mask function.  However, all of the examples of notch filter
mask functions discussed in this paper---except for the circular masks---have
periodic Fourier transforms.
The product of two such functions is another periodic notch filter
function.  For example, one notch filter mask
is
\begin{equation}
\hat M_{binary}(x,y)=\left\{
\begin{array}{ll}
0  & \mbox{where $|y-n \lambda/D| <   \hat M_{notch}(x)\lambda/(2D)$} \\
   & \qquad \mbox{or $|x -m \lambda/D| <   \hat M_{notch}(x)\lambda/(2D)$} \\
1  & \mbox{elsewhere}.
\end{array}
\right. 
\label{eq:maskproduct}
\end{equation}
In such a product, the bandwidths of the component masks add in each
direction separately. 

\begin{figure}
\epsscale{0.8}
\plotone{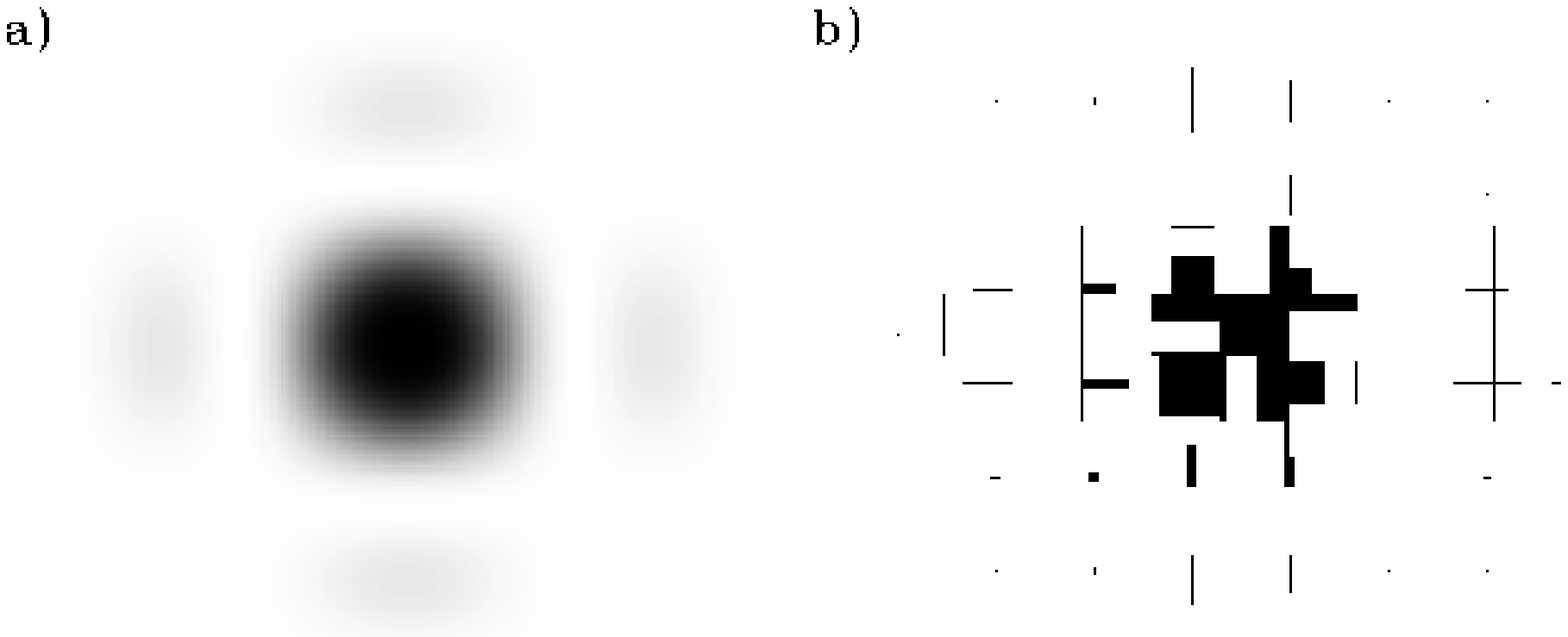}
\figcaption{Close-up view of the center of a $(1-{\rm sinc}^2 x)(1- {\rm sinc}^2 y)$ mask,
a band-limited mask which offers nearly as much search area as a Gaussian spot.
(a) Graded version (b) Binary notch filter version.
\label{fig:spotmask}}
\end{figure}

One may also produce a notch filter mask by combining the complements
of periodic notch filter masks.
For example,
\begin{equation}
\hat M_{binary}(x,y)=\left\{
\begin{array}{ll}
1  & \mbox{where $|y-n \lambda/D| >  (1-\hat M_{notch}(x))\lambda/(2D)$} \\
   & \qquad \mbox{or $|x -m \lambda/D| >   (1-\hat M_{notch}(x))\lambda/(2D)$} \\
0  & \mbox{elsewhere}.
\end{array}
\right. 
\label{eq:maskproduct2}
\end{equation}
Figure~\ref{fig:spotmask} shows a close up of a
mask with amplitude transmissivity
$(1-{\rm sinc}^2 x)(1- {\rm sinc}^2 y)$, and a binary
notch-filter version of this mask created by combining
the complements of two mask functions of the form $(1-{\rm sinc}^2)$.
This band-limited mask has a search area which closely
resembles that of a common mask which is not
band-limited---the Gaussian spot.
This mask and the one in Equation~\ref{eq:maskproduct}
have bandwidth in both the $x$ and $y$ directions.

As a third example, we can combine masks with 
sampled versions of the uniform mask, $\hat M(x)=C$, a constant.
Let
\begin{mathletters}
\begin{eqnarray}
\hat M_{T}(x) &=&  \hat P(x) * \Delta x \sum_n \delta(x-n \Delta x) \\
M_{T}(u) &=& P(u) \sum_n \delta(u-n/\Delta x) ,
\label{eq:dg}
\end{eqnarray}
\end{mathletters}
where $\int_{-\infty}^{\infty} \hat P(x) \, dx = C$.
We can multiply a notch filter mask function by $\hat M_{T}(x)$
and obtain another notch filter mask.
If we choose $\Delta x = \lambda/D$, and
$\hat P(x)=(D/\lambda)\Pi(x D/(C \lambda)$,
then the new mask will look just like the old one,
only painted with black stripes
of width $C \lambda/D$, spaced by $\lambda/D$
(Figure~\ref{fig:linearmasks}e), which may run in any direction.
Since $\hat M_{T}(x)$ diffracts some light outside the Lyot stop,
the intensity throughput of this new $\hat M_{notch}(x)$
will be reduced---by a factor of $C^2$.
Combining binary masks and these striped masks may make it
possible to design a range of masks which require no supportive
substrate.

\section{MASK ERRORS}
\label{sec:errors}

Consider a binary mask like the one shown in Figure~\ref{fig:linearmasks}d,
constructed from rectangles of opaque material, of width $\lambda/D$.  How
sensitive is the coronagraph to errors in the construction of
this mask?   What if one of these rectangles, in the center of the
mask, were accidentally made too short by an amount $h \lambda/D$, where
$h << 1$? 

A missing rectangle of material---or an extra rectangle of material---would
act like a tophat mask, diffracting light around the second pupil plane.
A tophat mask is not band-limited and it has a power spectrum that 
falls off quite slowly with spatial frequency.
Therefore it affords only modest cancellation of light in the
center of the second pupil plane. 

A tophat mask of width $\lambda/D$ and length $h \lambda/D$
produces a diffraction pattern with most of its power in a zone with
dimensions $D/\lambda$ by $D/ (h \lambda)$.  The intensity in this
illuminated region is proportional to $h$, but only a fraction
$\sim h$ of the illuminated region falls in the center of the Lyot stop.
In this portion of the illuminated region which falls in the center of the
Lyot stop, the field is roughly uniform, but attenuated by roughly a
factor of 2; the intensity is attenuated by a factor of four.  
Therefore, the final image will acquire an extra image of a point source
in the center, with fractional intensity $\sim 0.25 h^2$.

We can easily tolerate leakages of $\sim 10^{-7.5}$ of the starlight falling
in the center of the final image plane.  If we are to avoid leaks of greater
than this magnitude, we must avoid making the rectangles too short,
unless $0.25 h^2 \lesssim 10^{-7.5}$, or $h \lesssim 1/3000$.
For a telescope with focal ratio $f$, this tolerance becomes
$\lambda f/3000$, or typically $\sim 0.02$~ $\mu$m, for
$\lambda=1.0$~$\mu$m, $f=60$.

If the error is not in the center of the mask, but in the search area,
we can tolerate less leakage intensity, but the light falling on the hole
will be diminished in intensity by a similar amount,
so the requirement on the size of the hole remains about the same.
A hole far from the center, outside the search area, say at $100 \lambda/D$,
need only have $h < $ a few percent, since the wings of the stellar image
that fall on it are typically four or more orders of magnitude
weaker in intensity than the core of the stellar image. 

The tolerances for binary mask construction given here
fall within the reach of standard nanofabrication techniques.
Mask defects acquired during a mission may yield to compensation by 
the active optics a planet-finding coronagraph will
require to correct wavefront errors throughout the system.
If the hole in the mask described above had dimensions $\lambda/D$ by
$\lambda/D$, but had an intensity depth of only $f$, then
the hole would cause a fractional stellar
leakage of $\sim 0.25 f$, as opposed to $\sim 0.25 h^2$;
the shapes of binary masks are much less sensitive
to errors than the intensity transmissivities of graded masks.

\section{A WORKED EXAMPLE}
\label{sec:example}

To further illustrate the use of a notch-filter mask,
let us design some notch filter masks for a circular 4m TPF
coronagraph (see, e.g., Brown et al. 2002).  We will assume a bandpass from
$\lambda_{min}=0.66$~$\mu$m to $\lambda_{max}=1.0$~$\mu$m,
and a mask construction tolerance of 20 nm.
Rather than describing the optics in terms of the
dimensionless diffraction scale, $\lambda/D$, we will use the
phisical size of the diffraction scale in the focal plane,
$\lambda f$, where $f$ is the focal ratio of the telescope.
The search problem and the characterization problem call for
different specialized image masks, based on different
band-limited functions.  We will design a search mask---a 
linear mask rather than a circular mask
for the reasons enumerated in Section~\ref{sec:circular}.

If we choose a Lyot stop that works at $\lambda_{max}$, the
coronagraph will provide equal or better contrast over the whole band.
The half power point of the mask, $\theta_{1/2}$, is 
angle from the optical axis where
$\left(\hat M_{BL}(\theta_{1/2})\right)^2=0.5$;
we will choose $\theta_{1/2} = 3 \lambda_{max}/D$ or 150~mas.
This mask will enable us to search for planets as close as
$\sim 1$ AU projected distance from a star at 6.5 parsecs.
Searching for planets calls for a mask based on a band-limited
function with small wings, providing a large search area where
the planet image will be relatively
unattenuated by the mask.  A suitable one with
the prescribed half-power point has the
form $\hat M_{BL}=1 - {\rm sinc}^2(\pi x \epsilon / (\lambda_{max} f))$
with $\epsilon=0.4$ at $\lambda_{max}$, i.e.,
$\hat M=1 - {\rm sinc}^2(x/ (1.76 \lambda_{max} f))$.
Since the primary is circular, the shape of the Lyot stop will be
the overlap region of two unit circles whose centers are separated
by $\epsilon$, as depicted in Figure~4d of \citet{kuch02}.

To estimate the stellar leak due to pointing error and
the finite size of the stellar disk, we may re-write
Equation~16 in \citet{kuch02} so that it applies to
any linear mask which is roughly quartic inside its
half-power point.
The fraction of the starlight that
leaks through the mask, $L$, is
\begin{equation}
L \approx { {\theta_{\star}^4
+ 48 (\Delta \theta)^2 \theta_{\star}^2
+ 128 (\Delta \theta)^4 } \over {129 \ \theta_{1/2}^4}},
\end{equation}
where $\theta_{\star}$ is the angular diameter of the star,
$\Delta \theta$ is the pointing error in the $x$ direction,
and  is the half power point of the mask.
This equation also applies to nulling interferometers with
quartic nulls.  However, the half power point of a nulling
interferometer's fringe depends on wavelength.

When the pointing error is somewhat larger than the
typical stellar diameter ($\theta_{\star}=1.43$~mas for a G star at 6.5 pc),
the pointing error dominates the leak.
If the pointing errors, $\Delta \theta$, are
distributed in a Gaussian distribution about $\Delta \theta= 0$, 
the mean pointing-error-dominated leak is
\begin{equation}
L \approx 3 \left( { {\sigma_{\Delta \theta}} \over {\theta_{1/2}}}\right)^4,
\qquad \hbox{for $\sigma_{\Delta \theta} > \theta_{\star}$,}
\end{equation}
where $\sigma_{\Delta \theta}$ is the standard deviation of
the distribution.  If we require a mean leakage of
$\sim 3 \times 10^{-8}$ of the starlight falling in the
center of the final image, the pointing error tolerance becomes
$\sigma_{\Delta \theta} \le \theta_{1/2}/100$, or 1.5~mas.

This leakage due to pointing error can easily be
suppressed to the $10^{-10}$ level in the search area given a
weakly apodized Lyot stop.  For example, an apodization function of the form
$\cos{\pi u \lambda/((1-\epsilon/2)D)}$ provides 2.5 orders of
magnitude of suppression at $3 \lambda/D$.
The total throughput of the coronagraph would be
$1-\epsilon=0.6$ without the Lyot stop apodization; with
the apodization realized as a binary mask, it is 0.358.
This apodization is workable, but not optimal; further work on
choosing matched pupil and image stops could improve the
overall system performance.

Clearly, the leakage due to pointing errors quickly shrinks for planet-finding
coronagraphs with larger inner working distances, like the
proposed Eclipse mission \citep{trau02a}.  For this 2~m class
telescope with inner working distance $\sim 300$ mas, a
pointing error of $\sigma_{\Delta \theta}=3$~mas suffices to match
the above performance.
Likewise, if the pointing errors and other low-spatial-frequency errors
could be controlled below the levels assumed here,
these coronagraph designs could operate at smaller inner working distances. 

We will realize the mask as a binary notch filter mask like
the one in Figure~\ref{fig:linearmasks}d.  
In Section~\ref{sec:errors}, we showed that the lengths of the
bars in this mask must be accurate to $\lambda f/3000$.
Since a hole of a given physical size subtends a larger fraction
of $\lambda/D$ at smaller $\lambda$, this tolerance
applies at $\lambda_{min}$.  In other words, if we can manufacture
a mask accurate to 20~nm, we require a focal ratio $f=90$.
The strips must have maximum width $< \lambda_{min} f = 59.4 \mu$m.
Explicitly, the mask function would be:
\begin{equation}
\hat M_{binary}(x,y)=\left\{
\begin{array}{ll}
0  & \mbox{where $|y-n \lambda_{min} f | <  {{\lambda_{min} f }\over{ 2}} \hat M_{notch}(x)$} \\
1  & \mbox{elsewhere}.
\end{array}
\right. 
\end{equation}
where 
\begin{equation}
M_{notch}(x) = \sum_n \left( 1-{\rm sinc}^2 \left({{\pi \epsilon (n+\zeta) \lambda_{min}} \over { \lambda_{max}}}\right) \right)
 \Pi\left({{x} \over {\lambda_{min} f}}-(n+\zeta) \right) - \hat M_0.
\end{equation}
Table~1 shows that our mask has $\hat M_0 = 0.0054400$ and
$\zeta = 0.1017713$.  

In the absence of noise, analyzing the data from this
coronagraph is trivial.  A planet at angle $\theta_{p}$ from
the optical axis is simply attenuated by the low spatial frequency
components of the notch filter mask, i.e., 
$\approx \left(1 - {\rm sinc}^2(\theta_{p} / 90 {\rm mas} ) \right)^2$.
The shape of the PSF is set entirely by the Lyot stop;
it is the squared
absolute value of Fourier conjugate of the
Lyot stop amplitude transmissivity.

\section{SIMULATION OF CORONAGRAPH PERFORMANCE}

We numerically simulated the performance of this notch-filter
coronagraph design by following the Fraunhofer propagation
of light through a coronagraph using fast Fourier transforms.
\citet{naka94}, \citet{stah95} and \citet{siva01} have used
this technique to model the performance of ground-based coronagraphs.
We simulated the broad-band performance by running
the monochromatic simulation 10 times over a range of
wavelengths from 0.66 to 1.0~$\mu$m and averaged the
output images weighted by the stellar flux, assuming
a Rayleigh-Jeans law star and planet.
The noise representations remained the same from
wavelength to wavelength---scaled appropriately to model
pathlength errors rather than phase errors and to reflect
the change in the diffraction scale.

We used a 1024 by 1024 grid with resolution $\lambda_{min}/(2D)$.
In this representation, errors in the shape of the image mask become
variations in the mask amplitude transmissivity.
For example, if a bar in the binary mask were too long by 
$h \lambda/D$, the values of four adjacent elements in
the mask amplitude transmissivity matrix would
increase by $h$.

We assumed a circular pupil and a Lyot stop with the shape of
two overlapping circles as described above
multiplied by $\cos{\pi u /((1-\epsilon/2) D))}$.
Seen from afar, the contrast between the Earth and the Sun
when the Earth is at maximum angular separation is
$2 \times 10^{-10}$ \citep{desm01}.  We used this contrast
level for the planets in our simulation.

\begin{figure}
\epsscale{0.8}
\plotone{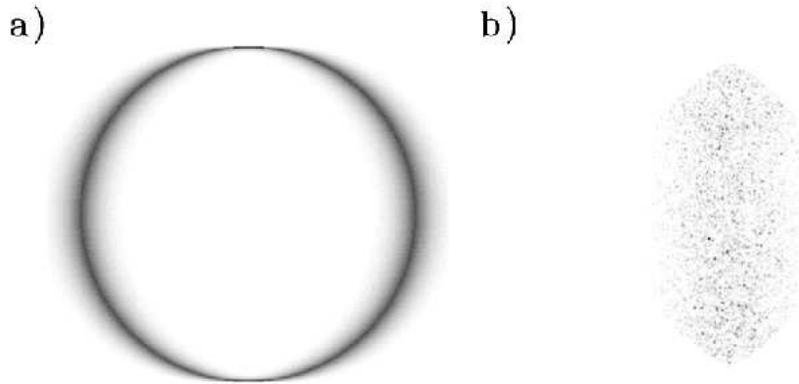}
\figcaption{Simulated monochromatic intensity in the Lyot plane
for a star and planet system imaged by the coronagraph described
in Section~\ref{sec:example}
a) before the Lyot stop
b) after the Lyot stop, amplified by a factor of $10^{9}$.
\label{fig:pupilplane}}
\end{figure}

Figure~\ref{fig:pupilplane} shows the intensity in the second pupil
plane before (a) and after (b) the Lyot stop
for a monochromatic simulation at 1.0~$\mu$m
of a system containing a star and a planet at $20 \lambda_{max}/D$.
The intensity after the Lyot stop has been multiplied by a factor of
$10^{9}$.  This figure represents a more accurate version
of the cartoon in Figure~\ref{fig:cartoon}.
Figure~\ref{fig:pupilplane}a shows how the diffracted light
falls within regions of width $\epsilon$ around the left and right edges
of the Lyot stop.  The notch filter mask adds further
illumination to the Lyot stop, farther off axis.  Our simulation
does not model these artifacts; the Lyot stop blocks them completely.
The planet adds a uniform background
illumination to the region inside the Lyot stop, though
the noise peaks due to wavefront errors and mask errors
dominate Figure~\ref{fig:pupilplane}b.

\begin{figure}
\epsscale{0.8}
\plotone{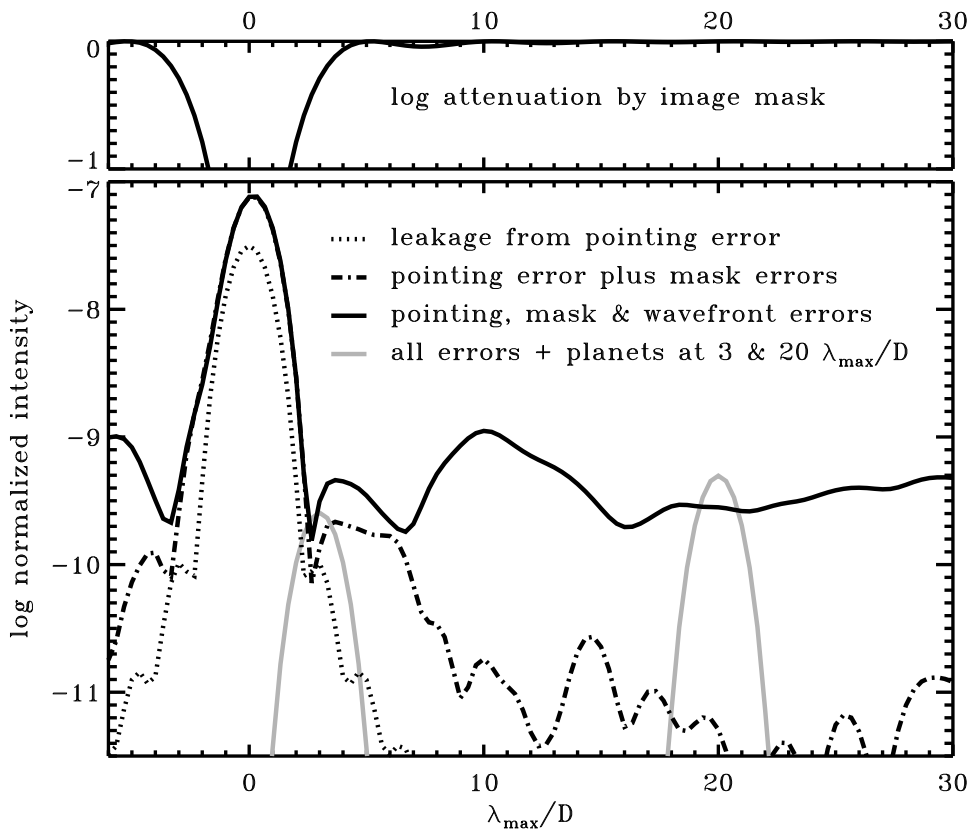}
\figcaption{Broadband simulation of images produced by the 
notch filter coronagraph design described in the text.
The dotted curve shows leakage due to pointing error.
The dot-dash curve adds errors to the lengths of the mask bars.
The solid curve adds amplitude and phase errors to the incoming wavefront.
The grey curves show the images of planets with relative flux
$2 \times 10^{-10}$ at 3 $\lambda_{max}/D$ and 20 $\lambda_{max}/D$.
The upper panel shows the attenuation caused by the
image mask.
\label{fig:simulation}}
\end{figure}

Figure~\ref{fig:simulation} shows a cut through the final
image plane along the $x$-axis.
The top panel of the figure shows a plot of 
$\log \left(1 - {\rm sinc}^2(\theta_{p} / 90 {\rm mas} ) \right)^2$.
This quantity is the
intensity attenuation supplied by the coronagraph, neglecting
any modification of the low spatial frequency components of
the mask that might occur in the construction of
a notch filter representation.

The lower panel of the figure shows the relative surface brightness of
several noise contributions to the image,
normalized to the peak of what the stellar image would be if the
image mask were removed.
The dotted curve shows the $10^{-7.5}$ leakage due to
pointing error which we prescribed.  The form of this curve is
simply the Fourier transform of 
the Lyot stop averaged over the band weighted by the stellar flux.
The dash-dot curve shows the consequence of adding
white noise to the lengths of the bars with r.m.s. 20 nm.
This noise concentrates near the optical axis in the final image plane
because it is multiplied by the image of the star that falls on the
image mask.  This phenomenon suggests that the tolerance
of the coronagraph to mask errors can be altered, and possibly
improved by manipulating the shape of the entrance pupil.

The solid black curve shows the consequence of adding white-noise
phase and amplitude errors to the incoming wavefront:
fractional amplitude errors of r.m.s. $10^{-3}$ over spatial frequencies
corresponding to the search area in the image plane (3--60 $\lambda/D$)
and phase errors due to pathlength errors of r.m.s 0.5~\AA\
over these frequencies.  A pathlength error of
0.5 \AA\ corresponds to an error in the figure of a mirror of 0.25 \AA.
Preliminary tests in the High Contrast Imaging Testbed at JPL
have demonstrated deformable mirror
wavefront control at this level \citep{trau02a, trau02b}.
Wavefront errors clearly
dominate mask errors and pointing errors, except within
a few diffraction widths of the optical axis.

The grey curves show the images of two planets
located  at $3 \lambda_{max}/D$ and $20 \lambda_{max}/D$, i.e.,
154 mas and 1026 mas---or 0.8~AU and 5.1~AU for a star 5 pc distant.
The planet at $3 \lambda_{max}/D$ is attenuated by a factor
of $0.5$ because it sits at the mask's half power point.
If $Q$ is the contrast between the planet's peak intensity
and the local scattered light background \citep{brow90},
the $3 \lambda/D$ planet
has $Q \approx 0.5$, and the $20 \lambda/D$ planet has
$Q \approx 1$.  Planets with $Q \sim 1$ can easily be
detected in a coronagraph using spectral deconvolution
techniques, for example, given sufficiently low photon
noise \citep{spar02}.

\section{CONCLUSION}

We have illustrated the use of notch filter functions to
generate several kinds of image masks which should be
relatively easy to manufacture.  We showed 
graded masks whose transmissivities are everywhere
greater than zero.  We showed binary image masks, which can be
cut or shaped from pieces of opaque material.
These binary masks can be manufactured to the tolerances necessary for
terrestrial planet finding using standard
nanofabrication techniques, and can potentially be
made self-supporting.  Our simulations of the performance
of a coronagraph outfitted with a binary notch filter mask
suggest that this technique could reveal
extrasolar planets similar in brightness to the Earth
around nearby stars, given foreseeable improvements in
wavefront control on a highly stable space platform.  

Binary notch filter masks combine many of
the advantages of binary pupil masks (ease of manufacture,
achromaticity, robustness) with the advantages of band-limited image
masks (large search area, and small inner working distance).  
Using binary pupil or image masks seems to inevitably
require stacking many copies of the same basic aperture shape;
\citet{kasd01} used this principle to
generate binary pupil masks; we have used it to 
generate binary image masks.  In \citet{kasd01}, the
high-spatial frequency artifacts of this stacking procedure
appear in the image plane directed away from a search sector.
In notch filter masks, the high-spatial frequency artifacts
are directed into the Lyot stop.

Ultimately, a space telescope for direct optical imaging of
extrasolar planets may incorporate more than one diffracted-light
management strategy.
Having a choice of different techniques
available will allow a mission to adapt to changing observing needs
as our understanding of high-contrast space telescopes improves and
the phenomenology of extrasolar planets unfolds.

\acknowledgements

We thank Chris Burrows, Wesley Traub and Dwight Moody for
helpful conversations.

This work was performed in part under contract with the Jet
Propulsion Laboratory (JPL) through the Michelson Fellowship
program funded by NASA as an element of the Planet Finder
Program.  It was funded in part by Ball Aerospace
Corporation via contract with JPL for TPF design
studies.  JPL is managed for NASA by the California
Institute of Technology.

\end{document}